\documentclass[reprint,
superscriptaddress,
shortbibliography,
amsmath,amssymb,
aps,
prr, 
floatfix 
]{revtex4-2}

\usepackage{xcolor}
\usepackage{diagbox}
\usepackage{graphicx}
\usepackage{bm}
\newcommand{\ve}[1]{\boldsymbol{#1}}

\begin{document}
\title{Circular dichroism in the photoelectron angular distribution of achiral molecules}

\author{Christian S. Kern}
\email[E-mail address: ]{christian.kern@uni-graz.at}
\affiliation{Institute of Physics, NAWI Graz, University of Graz, 8010 Graz, Austria}%

\author{Xiaosheng Yang}
\affiliation{Peter Gr\"unberg Institut (PGI-3), Forschungszentrum J\"ulich, 52425 J\"ulich, Germany}
\affiliation{J\"ulich Aachen Research Alliance (JARA), Fundamentals of Future Information Technology, 52425 J\"ulich, Germany}
\affiliation{Experimental Physics IV A, RWTH Aachen University, 52074 Aachen, Germany}

\author{Giovanni Zamborlini}
\affiliation{Peter Gr\"unberg Institut (PGI-6), Forschungszentrum J\"ulich, 52425 J\"ulich, Germany}
\affiliation{Institute of Physics, NAWI Graz, University of Graz, 8010 Graz, Austria}%

\author{Simone Mearini}
\affiliation{Peter Gr\"unberg Institut (PGI-6), Forschungszentrum J\"ulich, 52425 J\"ulich, Germany}

\author{Matteo Jugovac}
\affiliation{Peter Gr\"unberg Institut (PGI-6), Forschungszentrum J\"ulich, 52425 J\"ulich, Germany}

\author{Vitaliy Feyer}
\affiliation{Peter Gr\"unberg Institut (PGI-6), Forschungszentrum J\"ulich, 52425 J\"ulich, Germany}

\author{Umberto De Giovannini} 
\affiliation{Max Planck Institute for the Structure and Dynamics of Matter, 22761 Hamburg, Germany}
\affiliation{Dipartimento di Fisica e Chimica--Emilio Segr\`e, Universit\`a degli Studi di Palermo, 90123 Palermo, Italy}

\author{Angel Rubio}
\affiliation{Max Planck Institute for the Structure and Dynamics of Matter, 22761 Hamburg, Germany}
\affiliation{Initiative for Computational Catalysis (ICC), The Flatiron Institute, New York, New York 10010, USA}

\author{Serguei Soubatch}
\affiliation{Peter Gr\"unberg Institut (PGI-3), Forschungszentrum J\"ulich, 52425 J\"ulich, Germany}
\affiliation{J\"ulich Aachen Research Alliance (JARA), Fundamentals of Future Information Technology, 52425 J\"ulich, Germany}

\author{Michael G. Ramsey}
\affiliation{Institute of Physics, NAWI Graz, University of Graz, 8010 Graz, Austria}

\author{F. Stefan Tautz}
\email[E-mail address: ]{s.tautz@fz-juelich.de}
\affiliation{Peter Gr\"unberg Institut (PGI-3), Forschungszentrum J\"ulich, 52425 J\"ulich, Germany}
\affiliation{J\"ulich Aachen Research Alliance (JARA), Fundamentals of Future Information Technology, 52425 J\"ulich, Germany}
\affiliation{Experimental Physics IV A, RWTH Aachen University, 52074 Aachen, Germany}

\author{Peter Puschnig}
\email[E-mail address: ]{peter.puschnig@uni-graz.at}
\affiliation{Institute of Physics, NAWI Graz, University of Graz, 8010 Graz, Austria}


\begin{abstract}
Circular dichroism in the angular distribution (CDAD) is the effect that the angular intensity distribution of photoemitted electrons depends on the handedness of the incident circularly polarized light. The origin of CDAD can be manifold, including intrinsic properties of the system under study, such as chirality, spin-orbit interaction, or quantum-geometrical properties, but CDAD can also originate from final-state effects influenced by the experimental geometry. For example, CDAD has been reported for achiral organic  molecules at the interface to metallic substrates. For this latter case, we investigate two prototypical $\pi$-conjugated molecules, namely tetracene and pentacene, whose frontier orbitals have a similar shape but exhibit distinctly different symmetries. By comparing experimental CDAD momentum maps with simulations within time-dependent density functional theory, we show how the final state of the photoelectron must be regarded as the source of the CDAD in such otherwise achiral and quantum-geometrically trivial systems. We gain additional insight into the mechanism by employing a simple scattering model for the final state, which allows us to decompose the CDAD signal into partial wave contributions.
\end{abstract}

\maketitle

\section{\label{sec:intro}Introduction}
In optics, circular dichroism (CD) encompasses a set of spectroscopic techniques that are utilized to probe the differential absorption of circularly polarized light in matter. This subtle effect arises only in chiral materials and originates from the coupling between electric and magnetic transition dipoles~\cite{Caldwell1971}. CD spectroscopy reveals valuable information about the optical activity in chiral molecules~\cite{Woody2012} and is an important tool to probe the chemical environment of coordination complexes~\cite{Kirschner1969} or the conformation in large (bio-)molecules~\cite{Kelly2005, Berova2007, Pescitelli2011}. Even for achiral molecules, a magnetic transition moment, and thus a CD effect, may be induced by neighboring chiral hosts~\cite{Allenmark2003} or by applying a magnetic field~\cite{Buckingham1966}. CD is also observed for achiral molecules in the infrared regime~\cite{Deutsche1968, Hsu1973, Holzwarth1974}, which requires a non-zero magnetic dipole moment in vibrational transitions, an effect beyond the Born-Oppenheimer approximation, originating from moving charge distributions~\cite{Craig1978, Stephens1985}.

For photoemission triggered by circularly polarized radiation, photoelectron circular dichroism (PECD) from chiral, unoriented molecules had been theoretically predicted by Ritchie and Cherepkov in the 1970s~\cite{Ritchie1975, Ritchie1976, Ritchie1976a, Cherepkov1983} but was not experimentally confirmed until the turn of the millennium~\cite{Boewering2001}. Since then, PECD is a growing technique that is used to probe structural and electronic properties in various chiral materials~\cite{Jahnke2004,Fasel2004, Powis2008, Janssen2014, Ferrari2015,Turchini2017}, induced chirality~\cite{Mugarza2010, Contini2012, Contini2013,Rouquet2023}, chemical reactions~\cite{Baykusheva2019,Svoboda2022}, and has recently been extended to the time domain~\cite{Comby2016,Beaulieu2016,Tikhonov2022,Facciala2023,Monti2024,Wanie2024}. In contrast to optical CD spectroscopy, the PECD effect arises from electronic dipole transitions only and is therefore orders of magnitude more sensitive.

When measuring photoelectrons additionally with angular resolution (ARPES), circular dichroism in the angular distribution (CDAD) emerges in momentum maps of \textit{both} chiral and achiral structures~\cite{Ritchie1975,Ritchie1976, Ritchie1976a,Cherepkov1982,Cherepkov1983} and was first observed in a pump-probe experiment for an excited state in NO~\cite{Appling1986}. In this case, exciting the molecule prior to the photoexcitation step fulfilled the requirement of an \textit{aligned} initial state for CDAD~\cite{Dubs1985,Dubs1985a,Dubs1986}, which was also achieved in the case of graphite~\cite{Schoenhense1991} or small molecules~\cite{Westphal1989,Westphal1991} adsorbed on single-crystal surfaces. Given an oriented, but not necessarily chiral, initial state, certain experimental geometries, as defined by incidence and emission angles, may give rise to a dichroism in the angular-momentum-dependent photocurrent~\cite{Dubs1985}. As illustrated in the seminal work by Sch{\"o}nhense~\cite{Schoenhense1990}, this can be understood as a \textit{final-state} effect.

In addition to final-state effects due to the handedness of the experimental geometry, circular dichroism in photoemission is also associated with additional quantum numbers. For instance, when also measuring the spin of photoelectrons emitted by circularly polarized light, the degree of spin polarization depends on the helicity of the incident light.
Excluding spin-polarized initial states in magnetic materials~\cite{Baenninger1970,Henk1996}, the spin polarization of photoelectrons is then a relativistic effect predicted by Fano~\cite{Fano1969} and originates from spin-orbit coupling in either the photoemission matrix element or the final state~\cite{Kirschner1981,Heinzmann2012,Hartung2016,Vasilyev2020,Artemyev2024}.

In solids with strong spin-orbit coupling and spin-momentum locking, circular dichroism has also been shown to qualitatively follow the momentum-resolved spin texture~\cite{Zhang2014, Razzoli2017}. More generally, in two-dimensional systems and at crystal surfaces, linear or circular dichroism in ARPES has been linked to \emph{pseudospin} quantum numbers---such as sublattice, valley, or layer degrees of freedom---resolved via the (local) orbital angular momentum (OAM) distribution across the Brillouin zone~\cite{Park2012a,Cho2018,Beaulieu2020,Yen2024,Brinkman2024,Schusser2024,Sidilkover2025}. Those OAM or pseudospin characteristics may then be used to map topological properties, such as the Berry curvature~\cite{Schueler2020,Schueler2020a,Schueler2022}, or even the full quantum geometric tensor, as suggested recently~\cite{Kang2025,Kim2025}. With notable exceptions~\cite{Erhardt2024}, such conclusions generally cannot be drawn from experiment alone; substantial theoretical input is required to disentangle final-state effects from intrinsic material properties. In some cases, this necessitates sophisticated treatments of photoelectron scattering in the final state or layer-dependent attenuation~\cite{Sidilkover2025}, whereas in others the initial-state properties dominate and a plane-wave final-state model suffices~\cite{Schueler2020,Schueler2022}. Consequently, whether and how ARPES can access quantities such as OAM is system dependent, and the generality of these approaches remains uncertain. To isolate the influence of the photoemission final state on CDAD, it would therefore be advantageous to study systems with trivial quantum geometry or trivial (pseudo) spin textures but well-defined initial states with respect to symmetry and angular-momentum quantum numbers.

Such properties are realized in organic molecular layers. For these systems, photoemission orbital tomography (POT)~\cite{Puschnig2009a,Woodruff2016,Puschnig2018} has provided valuable insights, including the determination of structural properties of molecular adsorbates on surfaces~\cite{Zamborlini2017,Hurdax2022}, the identification of chemical reaction intermediates~\cite{Yang2019,Cojocariu2021}, the  deconvolution of the electronic structure~\cite{Puschnig2011,Koller2007,Wiessner2013a,Koller2007,Wiessner2013a,Haags2025}, or the reconstruction of real space orbitals from experimental data~\cite{Lueftner2014b,Wiessner2014,Weiss2015,Graus2019,Jansen2020,Dinh2024}, to quote only a few examples. However, the commonly applied simplification of a plane-wave final state, which enables the direct connection between the measured photoelectron angular distribution (or momentum map), and the initial state's Fourier transform renders the CDAD effect inaccessible~\cite{Schoenhense1990,Bradshaw2015,Dauth2016a,Egger2019,Moser2023}.

\begin{figure}[h!]
  \includegraphics[width=7.5cm]{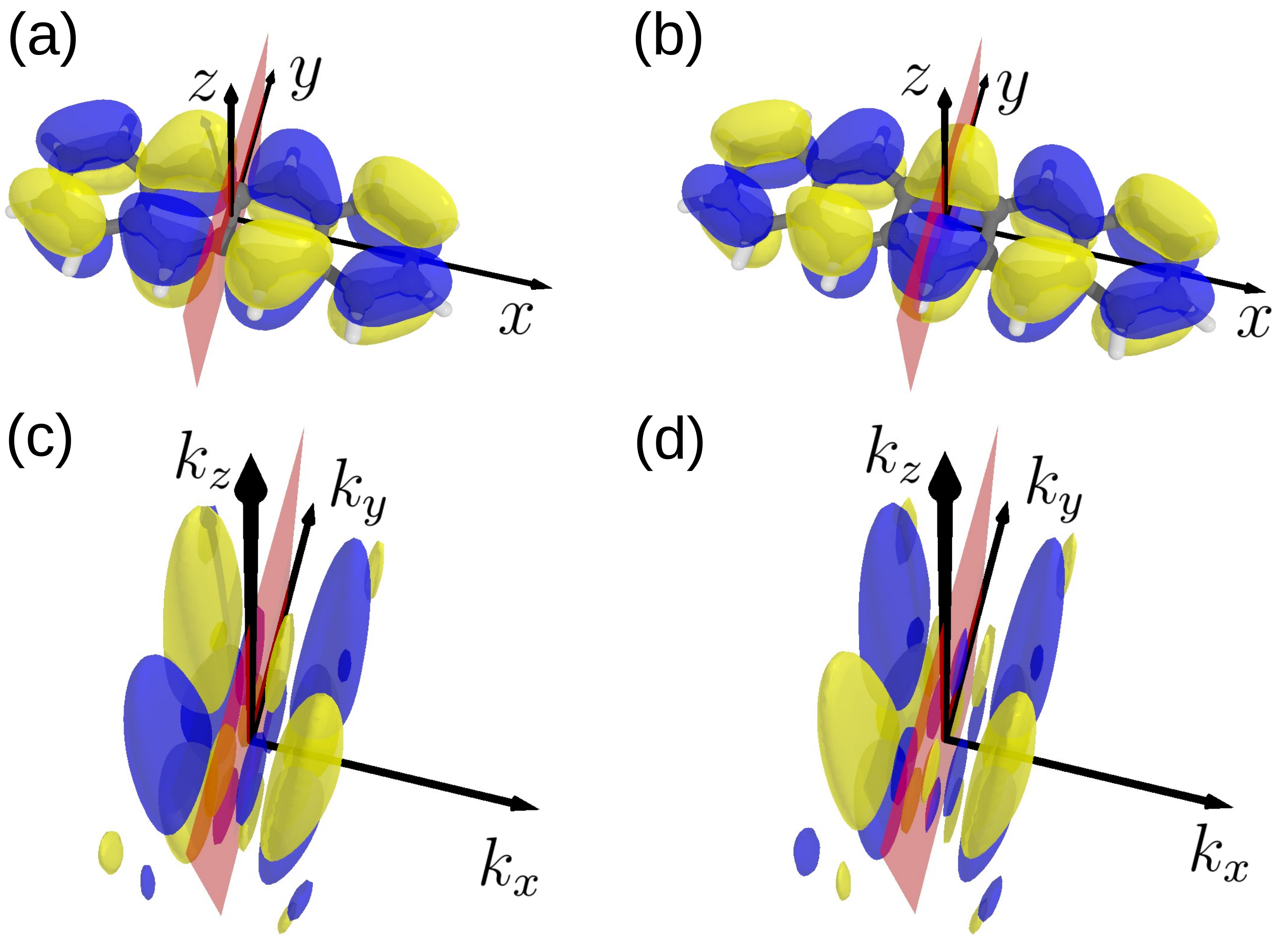}
  \caption{\label{fig:appx_orb} HOMOs of isolated tetracene (4A) (a, irreducible representation: $a_u$) and pentacene (5A) (b, irreducible representation: $b_{3g}$), the Fourier-transformed HOMOs of 4A [panel (c)] and 5A [panel (d)], and symmetry plane in red. The symmetry group of both 4A and 5A is $\mathrm D_\mathrm{2h}$ or $\frac{2}{m}\,\frac{2}{m}\,\frac{2}{m}$.}
\end{figure}

Apart from early studies with small molecules like CO or benzene~\cite{Westphal1989,Schoenhense1990}, experimental research on CDAD fueled by the full capabilities of modern momentum microscopy is still scarce. As an exception, dichroic momentum maps of the highest-occupied molecular orbital (HOMO) and formerly lowest-unoccupied molecular orbital (LUMO) of PTCDA on Ag(110) have been measured by Wie{\ss}ner \emph{et al.}~\cite{Wiessner2014}. In their work, the authors make the further claim that the phase symmetry of the initial state \textit{molecular} orbital can be determined from CDAD maps. More specifically, they proposed that if the real-space symmetry (and thus the group representations) of a molecule are known, one can deduce the phase symmetry of its wave functions from dichroic momentum space signatures, measured for two perpendicular light incidence directions. If true, this would allow the full quantum mechanical wave function to be determined experimentally~\cite{Wiessner2014}.

\begin{figure*}[htb]
  \includegraphics[width=0.8\textwidth]{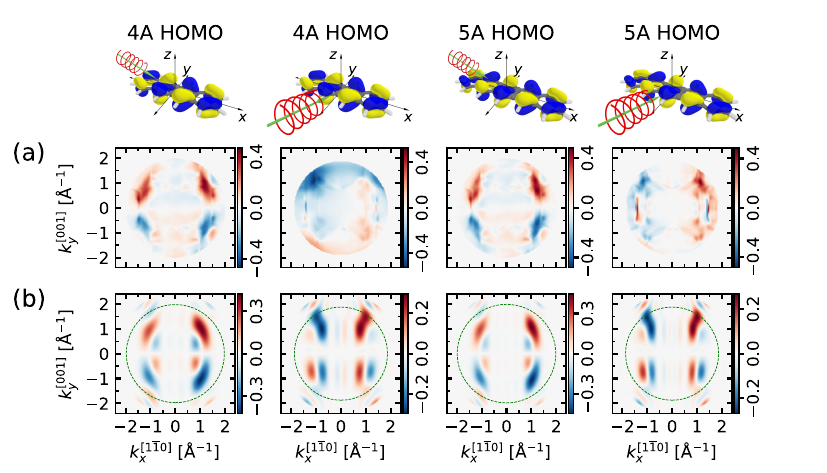}
  \caption{\label{fig:4A_5A_HOMO} CDAD maps for 4A and 5A and two different incidence directions of the circularly polarized light as indicated in the schematics. (a) Experimental CDAD maps of the 4A HOMO (binding energy: 1.65~eV) with the plane of incident light along the long and short molecular axes, and 5A HOMO (binding energy: 1.45~eV) with the plane of incident light along the long and short molecular axes (from left to right) on Cu(110). (b) Corresponding maps as simulated from TDDFT for isolated molecules. The green, dashed circles indicate the respective experimental photoemission horizon. Note the different scales for the CDAD magnitude.}
\end{figure*}

In the present combined experimental and theoretical study, we focus on CDAD in achiral molecular systems, also to test this hypothesis further. To this end, we study the influence of the final state for achiral systems with well-defined initial states of similar angular-momentum character but different phase symmetries. To achieve this, we have chosen two achiral representatives of the acene family, namely, tetracene (4A) and pentacene (5A). Both molecules form oriented monolayers on the Cu(110) and Ag(110) substrates and have already been investigated with POT using linearly polarized light sources~\cite{Yang2019a,Ules2014}. Importantly, while the HOMOs of 4A and 5A have a very similar electron distribution, they are of different symmetry (i.e., belong to different irreducible representations of the D\textsubscript{2h} point group), as illustrated in Figure~\ref{fig:appx_orb} in real and momentum space. Following the reasoning of Wie{\ss}ner \emph{et~al.}~\cite{Wiessner2014}, this difference in symmetry should result in distinctly different patterns in their CDAD momentum maps. We test this hypothesis by conducting ARPES experiments with circularly polarized light and by performing time-dependent density functional calculations using the surface flux method to simulate the angular emission patterns~\cite{Wopperer2017}. While we find excellent agreement between the measured data and the \emph{ab initio} simulations, we cannot, however, confirm a dependence of the CDAD momentum patterns on the orbital symmetry, thereby challenging the conclusions of the earlier work by Wie{\ss}ner \emph{et al.}~\cite{Wiessner2014}. Importantly, simulated CDAD momentum maps from an additional scattered-wave approach allow us to trace the observed CDAD effect back to the final-state description of photoemission. As a further result, we show how the very sensitive CDAD effect even allows to extract minor contributions of angular-momentum functions to the molecular orbitals. 

\section{\label{sec:results}Results}
\subsection{Experimental CDAD maps}
In our experimental approach, we deposit 4A and 5A on the Cu(110) surface, which results in well-ordered monolayers with a single molecular orientation~\cite{Chen2003}. Both adsorb with their aromatic planes parallel to the surface and with their long molecular axes parallel to the [1$\overline{1}$0] azimuthal direction. Photoemission tomography experiments were performed at the NanoESCA beamline of the Elettra synchrotron, Trieste, using circularly polarized light with an incidence angle of 65$^{\circ}$ with respect to the surface normal and a photon energy of 35~eV. In this photon energy regime, matrix element effects are dominating, while at higher photon energies electron diffraction becomes important for ARPES and consequently for the CDAD~\cite{Daimon1993}. Experimental details are presented in Appendix A.

In Figure~\ref{fig:4A_5A_HOMO}(a) we compare the photoemission results for the HOMOs of 4A and 5A. The single molecular orientation allows experiments to be performed in two principal geometries: with the incident light plane either \textit{parallel} (first and third columns) or \textit{perpendicular} (second and fourth columns) to the  long molecular axis, as sketched in the top row of Figure~\ref{fig:4A_5A_HOMO}. Here, the green vectors indicate the incident light direction and the red spirals the rotating light polarization vector of the right-handed helicity (not to scale). Note that throughout this paper, we use the ``quantum mechanics'' convention of handedness, such that for right-handed circularly polarized light the electric field vector rotates clockwise when looking in the direction of propagation, and we define the photoemission CDAD signal as
\begin{align}
  \label{eq:cdad_definition}
  \mathcal{I}_{\mathrm{CDAD}}(k_x, k_y, E_{\mathrm{kin}}) = \frac{\mathcal{I}_{\mathrm{RH}}}{\max\{\mathcal{I}_{\mathrm{RH}}\}} - 
  \frac{\mathcal{I}_{\mathrm{LH}}}{\max\{\mathcal{I}_{\mathrm{LH}}\}},
\end{align}
with $\mathcal{I}_{\mathrm{RH/LH}}=\mathcal{I}_{\mathrm{RH/LH}}(k_x, k_y, E_{\mathrm{kin}})$ denoting the photoemission intensities obtained with right-handed and left-handed circularly polarized light, respectively, and $\max\{\mathcal{I}_{\mathrm{RH/LH}}\}$ is the maximum value of the respective photoemission intensity in the ($k_x$, $k_y$) plane. 

Focusing first on the experimental results for the parallel incidence direction, we notice that the CDAD momentum patterns for the 4A HOMO [Figure~\ref{fig:4A_5A_HOMO}(a), first column] and the 5A HOMO [Figure~\ref{fig:4A_5A_HOMO}(a), third column] have the same symmetry, which is remarkable given the fact that the  HOMOs of isolated 4A and 5A belong to different irreducible representations $a_u$ and $b_{3g}$, respectively. For instance, while 4A's HOMO is antisymmetric, 5A's HOMO is symmetric with respect to reflection at the ($y,z$) plane (see also Figure~\ref{fig:appx_orb}). We remark that this finding conclusively disproves the claim of ref.~\cite{Wiessner2014} by counterexample, namely, by identifying orbitals with \emph{different} symmetries that yet lead to dichroic momentum maps with \emph{identical} symmetries. In this context  it is also noteworthy  that 4A and 5A (studied by us) and PTCDA (studied by \citet{Wiessner2014}) are each adsorbed on comparable low-index coinage metal surfaces, which induce charge transfer into the adsorbed molecules and thereby fill the respective LUMOs. The HOMO of 4A and the HOMO of PTCDA  share the a\textsubscript{u} symmetry, while the HOMO of 5A and the LUMO of PTCDA share the b\textsubscript{2g} symmetry. Despite belonging to different symmetry groups, the HOMO of 5A and the HOMO of PTCDA, however, share the same symmetry pattern of the CDAD maps, for both light incidence directions. We, therefore, seek an explanation of the CDAD effect in the \textit{final} state of the photoemission process~\cite{Schoenhense1990}. Given that the HOMO does not strongly hybridize with the substrate, we can hope to capture the essential effects already within simulations of isolated molecules.

\subsection{Simulated maps from TDDFT}
Thus, as a first strategy, we simulate the photoemission final state in a bias-free approach within TDDFT~\cite{DeGiovannini2012,DeGiovannini2013,Wopperer2017,DeGiovannini2017} using the Octopus code~\cite{Marques2003, Castro2006, Andrade2015, TancogneDejean2020}. Employing the t-SURFF method originally developed for the time-dependent Schr{\"o}dinger equation~\cite{Tao2012, Scrinzi2012}, here the effective Kohn-Sham system is propagated in time, and photoemission is triggered by a time-dependent vector potential coupled to the electrons' momenta. Provided that the real-space simulation box for the molecule is sufficiently large and absorbing boundaries are applied~\cite{DeGiovannini2015}, photoelectrons can be numerically recorded with angular and kinetic energy resolution. In this way, no explicit final state needs to be assumed and the interaction of the outgoing photoelectrons with each other and with the remaining system are naturally accounted for. The main approximation then consists in the treatment of the exchange-correlation kernel; here, we choose the commonly applied adiabatic local density approximation~\cite{Dirac1930, Perdew1981, Yabana1996}. This method has already been successfully used to simulate CDAD maps for organic molecules including 4A~\cite{Wopperer2017} and for 2D materials~\cite{Schueler2020}. The numerical settings of our TDDFT simulations are detailed in Appendix~\ref{sec:appx_tddft}.

The TDDFT CDAD maps of 4A and 5A for the parallel incidence direction are shown in the first and third columns of Figure~\ref{fig:4A_5A_HOMO}~(b), respectively. For both molecules, the main features at $k_x=\pm 1.3~\textup{\AA}^{-1}$ agree very well with experiment, both in shape and in position. Additionally, the simulations show an alternating pattern of minor features inside and outside the main lobes, where the latter is not clearly seen in the experiment due to the experimentally limited $\ve k_{||}$ horizon, which results from the limited field of view of the detector ($\sim$90 \% of the theoretical $\ve k_{||}$ range) and additional losses in the post processing in order to avoid artefacts at the rims. We have have indicated the effective experimental $\ve k_{||}$ horizon in the corresponding simulated maps by a green circle. 

For the other light incidence direction (i.e., perpendicular to the long molecular axis, depicted in columns two and four of Figure~\ref{fig:4A_5A_HOMO}), we notice a change in the overall CDAD symmetry, which is now antisymmetric with respect to reflection at the ($k_y,k_z$) plane, compared to being antisymmetric with respect to the ($k_x,k_z$) plane for parallel incidence. Thus, in both cases the symmetry of the CDAD pattern follows the light incidence direction, suggesting it to be primarily caused by the light polarization. Again, the agreement between experiment and TDDFT is adequate, although the measurements in this case suffer from a lower relative intensity of the molecular versus the substrate emission features---the latter play a more dominant role in the momentum maps, which is especially pronounced in regions where both overlap in momentum space. This is the case around the $(k_x, k_y)=(\pm 1, \pm 1)~\textup{\AA}^{-1}$ points and presumably the reason why the outer features with opposite sign in the CDAD of the TDDFT maps are not observed in experiment. Judging from the TDDFT simulations, the CDAD effect for the molecular emission is more pronounced for the long-axis incidence direction ($\sim 35\%$) than for the short-axis incidence direction [$\sim 25\%$, see color scale in Figure~\ref{fig:4A_5A_HOMO}(b)], which also helps to explain why for the perpendicular incidence direction the measured momentum maps are weaker compared to the substrate emissions. In this regard, it should also be noted that the 4A HOMO resides at a slightly different binding energy (1.65~eV) than the 5A HOMO (1.45~eV), which may give rise to the different sign of the CDAD signal of the substrate $sp$-band features at $k_x\approx\pm~1.5~\textup{\AA}^{-1}$ for the 4A and 5A interfaces measured at different binding energies.

Summarizing the findings presented so far, we can state already at this stage that, similar to molecular photoemission with linearly polarized light, simulations of isolated molecules are able to capture the main features observed in experiment, however for CDAD if and only if final-state effects are properly taken into account. While TDDFT is capable of doing this, it is computationally very expensive and becomes prohibitive beyond isolated molecules or small unit cells of periodic systems and, above all, provides no direct insight into the microscopic origin of the CDAD effect in systems that are not dominated by band topology or (pseudo) spin polarization. In the following, we therefore aim for a more intuitive (and computationally less costly) description of circular dichroism in photoemission momentum maps.

\subsection{The scattered-wave approximation}
To describe circular dichroism in photoemission momentum maps in an intuitive and efficient way, we employ the independent atomic center approximation (IACA)~\cite{Grobman1978}. Here, the molecule's initial state $|i\rangle$ is expressed as a linear combination of local basis functions at each atom position $\ve R_a$,
\begin{align}
  \label{eq:initial_state}
  \langle \ve r |i \rangle = \sum_{a, n l m} c_{n l m}^a R_{n l}(r_a)Y_{l m}(\widehat{\ve r}_a),
\end{align}
with $\ve r_a = \ve r - \ve R_a$, radial functions $R_{n l}(r_a)$, the spherical harmonics $Y_{l m}(\widehat{\ve r}_a)$, $r_a$ the length of $\ve r_a $, and $\widehat{\ve r}_a$ the unit vector in the direction of $\ve r_a$. The final state $|f_{\ve k}\rangle$, characterized by a given wave vector $\ve k$, is also written as a linear combination of atom-centered contributions, each of which taking the form of a partial wave expansion
\begin{align}
  \label{eq:final_state}
  \langle \ve r |f_{\ve k} \rangle = 4\pi 
  \sum_{a, l m} \mathrm{e}^{\mathrm i \ve k\ve R_a}\mathrm{e}^{\mathrm i \sigma_{l}^Z} j_l(k\, r_a)
  \mathrm i^{l}\mathrm{Y}_{l m}^*(\widehat{\ve k}) 
  \mathrm{Y}_{l m}(\widehat{\ve r}_a).
\end{align}

Note that, for simplicity, we approximate the radial parts of the final state by spherical Bessel functions $j_l(k\, r_a)$, but take into account the phase shifts $\sigma_l^Z(k)$ between partial waves of different angular momenta by the Coulomb phase shifts
\begin{align}
  \label{eq:phase_shifts}
  \sigma_l^Z(k) = \arg \Gamma (l + 1 - \mathrm i Z /k).
\end{align}
Here, $\Gamma$ is the $\Gamma$ function, $k=|\ve k|$, $Z$ is an effective charge, and we use atomic units throughout this article. Such a scattered-wave ansatz has, in fact, a long history in the theoretical description of photoemission. For instance, it has been used to explain oscillations in the ARPES intensity in C$_{60}$~\cite{Hasegawa1998} and graphene~\cite{Kern2023}, as well as for circular dichroism in small molecules~\cite{Dubs1985a, Dubs1985, Westphal1989,Schoenhense1990} or solid-state systems~\cite{Moser2023}. Setting all $\sigma^Z_l=0$ would recover the plane-wave approximation.

\begin{figure}[htb]
  \includegraphics[width=8.4cm]{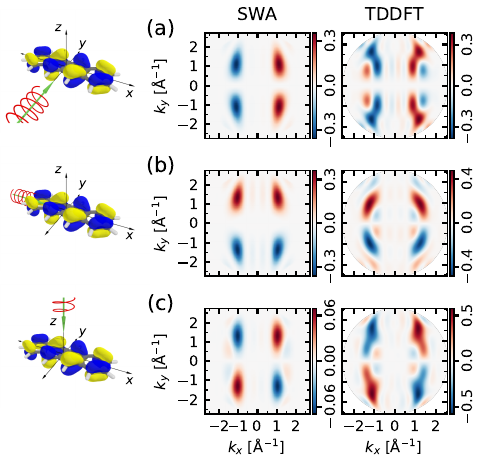}
  \caption{\label{fig:tddft_swa}SWA (left column) vs. TDDFT (right column) CDAD momentum maps of the 4A HOMO obtained for three different incidence directions of light, as indicated in the sketches. From top to bottom [(a)--(c)]: field configurations $\ve A_1$, $\ve A_2$, and $\ve A_3$, as described in the text.}
\end{figure}

With these ingredients, we can write the photoemission intensity in the scattered-wave approximation (SWA) and within Fermi's Golden rule as the sum over all transitions $|i\rangle \rightarrow |f_{\ve k}\rangle$, induced by the photon field $\ve A=A_0 \widehat{\ve A}$, as
\begin{align}
  \label{eq:me}
  \mathcal{I}(\ve k) = 2\pi \sum_i \left| 
  \langle f_{\ve k} | \ve{A} \cdot \nabla | i \rangle \right|^2
  \delta(\omega + E_i - E_f).
\end{align}
Here, we use the dipole approximation in the velocity gauge and, as an approximation to the spectral function, a Dirac delta function to enforce energy conservation with the photon energy $\omega$. As shown in Appendix~\ref{sec:appx_me}, the matrix element $\mathcal{M}(\ve k)=\langle f_{\ve k} | \ve{A} \cdot \nabla | i \rangle$ can then be expressed as
\begin{align}
  \label{eq:me_final}
  \mathcal{M}(\ve k) = 4 &\pi A_0 \sum_{a,nlm} c_{n l m}^a \mathrm{e}^{\mathrm i \ve k \ve R_a} \\
  \Bigg[&\mathrm{e}^{-\mathrm i \sigma_{l-1}^Z}\mathcal{R}_{a,n}^{(l-1)}
  \hspace{-3mm}\sum_{m'=-(l-1)}^{l-1}\hspace{-4mm}\mathrm{Y}_{(l-1) m'}(\widehat{\ve k})
  \alpha_{m,m'}^{(l,l- 1)}(\widehat{\ve A})\nonumber \\
  +&\mathrm{e}^{-\mathrm i \sigma_{l+1}^Z}\mathcal{R}_{a,n}^{(l+1)}
  \hspace{-2mm}\sum_{m'=-(l+1)}^{l+1}\hspace{-4mm}
  \mathrm{Y}_{(l+1) m'}(\widehat{\ve k})
  \alpha_{m,m'}^{(l,l+ 1)}(\widehat{\ve A})\Bigg]. \nonumber
\end{align}
In this equation, the $\mathcal{R}_{a,n}^{(l\pm1)}$ denote the radial integrals, defined in Eq.~\ref{eq:def_rad_me}, and  $\alpha_{m,m'}^{(l,l\pm 1)}(\widehat{\ve A})$ are the angular coefficients as functions of the unit field direction $\widehat{\ve A}$, defined in Eq.~\ref{eq:def_ang_coef}. The matrix element is now expressed in terms of two $(l\pm 1)$ channels, a well-known consequence of light-matter interaction in the dipole approximation, which we can exploit also here within the IACA by utilizing partial wave expansions for both the initial and the final states. In the following, we specify the values for the angular coefficients $\alpha_{m,m'}^{(l,l\pm 1)}(\widehat{\ve A})$ for the most relevant angular momentum channels in the initial and final states and discuss how this leads to the CDAD effect.

\begin{table}[htb]
\caption{\label{tab_ang_xz} Angular coefficients $\alpha_{m,m'}^{(l,l\pm 1)}(\widehat{\ve A}_1)$ for light polarization $\widehat{\ve A}_1=(\pm \mathrm i \hat{\ve x}+\hat{\ve z})/\sqrt{2}$ for combinations of initial states $|i\rangle=|l,m\rangle $ (columns) and final states $|f\rangle=|l\pm1,m'\rangle$ (rows) for both dipole-allowed channels. $\pm$  and $\mp$ signs reflect a sign change upon change of helicity.}
\begin{center} 
\setlength{\tabcolsep}{2pt}
\renewcommand{\arraystretch}{1.5}
\begin{tabular}{l|l|l|l|l|l|}
\cline{2-6}
& \diagbox{$|f\rangle$}{$|i\rangle$} & $|0,0\rangle$ & $|1,-1\rangle$ & $|1,0\rangle$ & $|1,1\rangle$\\ \cline{2-6}
\vspace{-.2cm} \\
$l-1$ \\
\hline \vspace{-.5cm} \\
\hline
 & $| 0 , 0 \rangle$ &    & $ \mp\frac{\mathrm{i}}{2} $ & $ \frac{\sqrt{2}}{2} $ & $ \pm \frac{\mathrm{i}}{2} $ \\ \cline{2-6}
\vspace{-.2cm} \\
$l+1$ \\
\hline \vspace{-.5cm} \\
\hline
 & $| 0 , 0 \rangle$ &    &    &    &    \\ \cline{2-6}
 & $| 1 , -1 \rangle$ & $ \pm \frac{\sqrt{3} \mathrm{i}}{6} $ &    &    &    \\ \cline{2-6}
 & $| 1 , 0 \rangle$ & $ - \frac{\sqrt{6}}{6} $ &    &    &    \\ \cline{2-6}
 & $| 1 , 1 \rangle$ & $ \mp\frac{\sqrt{3} \mathrm{i}}{6} $ &    &    &    \\ \cline{2-6}
 & $| 2 , -2 \rangle$ &    & $ \pm \frac{\sqrt{15} \mathrm{i}}{10} $ &    &    \\ \cline{2-6}
 & $| 2 , -1 \rangle$ &    & $ - \frac{\sqrt{15}}{10} $ & $ \pm \frac{\sqrt{30} \mathrm{i}}{20} $ &    \\ \cline{2-6}
 & $| 2 , 0 \rangle$ &    & $ \mp\frac{\sqrt{10} \mathrm{i}}{20} $ & $ - \frac{\sqrt{5}}{5} $ & $ \pm \frac{\sqrt{10} \mathrm{i}}{20} $ \\ \cline{2-6}
 & $| 2 , 1 \rangle$ &    &    & $ \mp\frac{\sqrt{30} \mathrm{i}}{20} $ & $ - \frac{\sqrt{15}}{10} $ \\ \cline{2-6}
 & $| 2 , 2 \rangle$ &    &    &    & $ \mp\frac{\sqrt{15} \mathrm{i}}{10} $ \\ \cline{2-6}
\end{tabular}
\end{center}
\end{table}

We start our analysis with the three prototypical cases of light incidence along the $\hat{\ve y}$, $\hat{\ve x}$, and $\hat{\ve z}$ Cartesian directions, respectively, as depicted in Figure~\ref{fig:tddft_swa}, with the corresponding vector potentials given by $\ve A_1=A_0(\pm \mathrm i \hat{\ve x} + \hat{\ve z})/\sqrt{2}$, $\ve A_2=A_0(\mp \mathrm i \hat{\ve y} + \hat{\ve z})/\sqrt{2}$, and $\ve A_3=A_0(\hat{\ve x} \pm \mathrm i  \hat{\ve y})/\sqrt{2}$. Here, the $\pm$ signs correspond to right- and left-handed cicular polarization, respectively. For the first case, $\ve A_1$, we present the values for $\alpha_{m,m'}^{(l,l\pm 1)}(\widehat{\ve A}_1)$ in Table~\ref{tab_ang_xz} for generic initial state orbitals $|l,m\rangle$ (columns) and all possible final states $|l\pm1,m'\rangle$. For the frontier orbitals of the acenes, clearly the most important contribution to the initial states is of carbon $p_z$ character ($|1, 0\rangle$), which results in $s$- and $d$-orbital character in the final state. Crucially, for the two $d$ contributions $|2, -1\rangle$ and $|2, 1\rangle$, we observe a sign reversal of $\alpha_{m,m'}^{(l,l\pm 1)}(\widehat{\ve A}_1)$ upon change of helicity, which will lead to a non vanishing CDAD signal.

This becomes most evident when making use of the real-valued spherical harmonics (also known as cubic or tesseral harmonics), defined as $s=\mathrm{Y}_{00}$, $d_{xz}=(\mathrm{Y}_{2-1}-\mathrm{Y}_{21})/\sqrt{2}$, $d_{z^2}=\mathrm{Y}_{20}$, etc. Inserting the values for $\alpha_{m,m'}^{(l,l\pm 1)}(\widehat{\ve A}_1)$ in Table~\ref{tab_ang_xz} into Eq.~\ref{eq:me_final}, we obtain the matrix element for photoemission from a collection (sum over $a$) of $p_z$ orbitals with principle quantum number $n$: 
\begin{align}
  \label{eq:me_xz}
  \mathcal{M}^{(\pm)}(\ve k) = &\frac{4\pi}{\sqrt{10}} A_0\sum_{a,n} c_{n 1 0}^a\mathrm{e}^{\mathrm{i} \ve k \ve R_a}
  \Bigg[\sqrt{5}\; \mathrm{e}^{-\mathrm i \sigma_{0}^Z} \mathcal{R}_{a,n}^{(0)} s(\widehat{\ve k}) \nonumber\\
  - &\mathrm{e}^{-\mathrm i \sigma_{2}^Z} \mathcal{R}_{a,n}^{(2)}\Big(\sqrt{2}d_{z^2}(\widehat{\ve k})\mp \mathrm i \sqrt{\frac{3}{2}} d_{xz}(\widehat{\ve k})\Big) \Bigg].
\end{align}
Here, the coherent summation in the matrix element leads to a mixing of the $s$, $d_{z^2}$, and $d_{xz}$ channels in the photoemission intensity. When evaluating the CDAD as defined in Eq.~\ref{eq:cdad_definition}, however, only the terms that change sign when changing the helicity will contribute. To illustrate the effect of such sign changes on the CDAD signal, we first consider photoemission from one atomic site and take a basis with only one radial function per $l$ channel (single $\zeta$), such that the sums over $a$ and $n$ in Eq.~\ref{eq:me_xz} collapse, leading to
\begin{align}
  \label{eq:cdad_xz}
  \mathcal{I}_{\mathrm{CDAD}}^{(a,n10)} \propto A_0^2 |c_{n 1 0}^{a}|^2
  \mathcal{R}^{(0)}_{a,n}\mathcal{R}^{(2)}_{a,n}
  s(\widehat{\ve k}) d_{xz}(\widehat{\ve k})
  \sin(\sigma_2^Z-\sigma_0^Z),
\end{align}
In essence, the CDAD is thus stemming from the fact that the photoelectron's final-state partial waves for the two dipole-allowed angular momentum channels experience a \textit{different} phase shift due to the atomic potential. The magnitude of the CDAD signal is directly proportional to the sine of this difference, while its angular distribution is given by the product of the angular functions $s(\widehat{\ve k})$ and $d_{xz}(\widehat{\ve k})$, resulting from the action of the dipole operator on initial-state wave functions of well-defined angular momentum.

It is important to keep in mind that for the above discussion, leading to Eq.~\ref{eq:cdad_xz}, we only considered photoemission from \emph{one} atomic site, which already suffices to explain the main mechanism for the CDAD effect, namely, the interference of different partial waves in the final state. However, when considering photoemission from \emph{all} atomic sites, in addition to the \emph{intra}atomic contributions ($a=a'$), also \emph{inter}atomic contributions ($a\neq a'$) must be considered~\cite{Sidilkover2025}. As detailed in Appendix~\ref{sec:appx_contributions}, we can split the total CDAD signal into these contributions in the following way,
\begin{align}
  \label{eq:cdad_xz_sum_a}
  \mathcal{I}_{\mathrm{CDAD}}^{(n10)} &\propto \Delta F(\widehat{\ve k}; Z) \bigg(\sum_a |\tilde{c}_{a}|^2 \\ \nonumber
  &+2 \sum_{a<a'} \tilde{c}_{a}\tilde{c}_{a'}^* \cos\left[\ve k (\ve R_a - \ve R_{a'})\right] \bigg),
\end{align}
where we have abbreviated $\Delta F(\widehat{\ve k}; Z) = A_0^2\mathcal{R}_{a,n}^{(0)}\mathcal{R}_{a,n}^{(2)}s(\widehat{\ve k}) d_{xz}(\widehat{\ve k}) \sin(\sigma_2^Z-\sigma_0^Z)$, as in Eq.~\ref{eq:cdad_xz}, and $\tilde{c}_a=c_{n10}^a$ for brevity. Note that the $\widehat{\ve k}$ dependence of $\Delta F(\widehat{\ve k}; Z)$ encodes effects of the intial state's local OAM, while its sign and magnitude are determined by interference of the dipole-allowed partial waves in the final state. The cosine term in the interatomic contributions gives rise to an additional $\ve k$ dependence~\cite{Sidilkover2025}, where the symmetry of the CDAD pattern can be traced back to the $\Delta F (\widehat{\ve k}; Z )$, while its internal structure is determined by the term in parenthesis which is closely related to the Fourier transform of the orbital.

In the following, we turn back to Eq.~\ref{eq:me_final} and present numerical results of the full SWA model for the 4A HOMO using the same  light incidence geometries $\ve A_1$, $\ve A_2$, and $\ve A_3$ as above. These simulations will then be compared to corresponding TDDFT calculations. For the SWA calculations, we use initial states obtained within density functional theory and the NWChem code~\cite{Apra2020}, employing the B3LYP functional~\cite{Becke1993,Stephens1994} and a cc-pvdz basis set. For the field configuration $\ve{A}_1$ and the HOMO of 4A, we present the SWA-calculated CDAD momentum map in the left panel of Figure~\ref{fig:tddft_swa}(a). Analyzing this result, we find an overall left-right antisymmetry of the main lobes around $k_x\approx \pm 1.2~\textup{\AA}^{-1}$, which have a mildly rounded shape. This shape stems from the fact that the CDAD maps result from a superposition of $s$ and $d_{xz}$ cubic harmonics, as opposed to $p_z$ for corresponding maps obtained with linearly polarized light (equivalent to  $\mathcal{I}_{\mathrm{RH}} + \mathcal{I}_{\mathrm{LH}}$ for circularly polarized light), which we show in Figure~\ref{fig:appx_linpol} in the Appendix for comparison. Furthermore, we note that the chemical environment for different atomic species is reflected in the parameter $Z$ of the scattering phase shift [Eq.~\ref{eq:phase_shifts}]. In our case of hydrocarbons, we found almost no contribution from hydrogen to the CDAD signal of the frontier orbitals, which was tested by switching off their contribution in the simulations. We also found only a scaling effect when changing the $Z$ parameter in a reasonable range and thus set $Z=0.5$ for carbon throughout this work. However, especially for structures containing both light and heavy elements, this effect might become important. 

For the case of light incidence along the long axis of the molecule, i.e., $\ve A_2=A_0(\mp \mathrm i \hat{\ve y} + \hat{\ve z})/\sqrt{2}$, we find a very similar mechanism as above, using the angular coefficients listed in Table~\ref{tab_ang_yz} of Appendix~\ref{appx:ang_tables}. In essence, simply the role of $x$ and $y$ are interchanged, leading to a $d_{yz}=\mathrm i(\mathrm{Y}_{2-1}+\mathrm{Y}_{21})/\sqrt{2}$ final-state symmetry instead of the $d_{xz}$ as before.  
It should be noted that precisely this example for the essential mechanism of CDAD, exemplified for a single $p_z$ orbital as the initial state, has already been discussed by Schönhense~\cite{Schoenhense1990}. In combination with the IACA and for the case of the 4A HOMO as a realistic initial state, we now observe an antisymmetry of the CDAD map with respect to the ($k_x,k_z$) plane, as depicted in the left panel of Figure~\ref{fig:tddft_swa}(b). The maps also exhibit the additional features that have already been discussed above for the  case of $\ve A_1$.

Finally, we turn to the third prototypical case with light incidence along the $z$ axis, despite the fact that for this normal-incidence geometry no measurements are feasible at the NanoESCA beamline endstation. However, as we will outline below, our simulation results for normal incidence are remarkable, since due to the axial symmetry of $p_z$ orbitals along $z$ one would expect no contributions from $p_z$ initial states to the CDAD signal, as confirmed by evaluating Eq.~\ref{eq:me_final} with the help of Tables~\ref{tab_ang_xy_RH} and~\ref{tab_ang_xy_LH} for this light incidence geometry. Consequently, when computing the 4A HOMO in a basis set that contains only $s$ and $p$ states (e.g., the 6-31G basis set), the CDAD signal vanishes for normal incidence as depicted in Figure~\ref{fig:appx_basis_sets}(b). In fact, to obtain a converged result for the CDAD maps, we found it necessary to use at least a cc-pvtz basis set, which additionally contains two $d$ and one $f$ function for carbon. In contrast, for grazing incidence ($\ve A_1$ and $\ve A_2$), we obtain already resonable results for the 6-31G basis set [see Figure~\ref{fig:appx_basis_sets}(b)]. From our basis set convergence studies and the above considerations, we can therefore conclude that with normal incidence, we probe the HOMO in terms of atomic orbital contributions that are not invariant with respect to rotations in the ($x,y$) plane. Since for a basis set with $l<2$ the CDAD is zero, this excludes $p_x$ and $p_y$ and suggests the contributions of $d$ states to the initial-state orbital to be the source of dichroism in this case. Indeed, when analyzing the SWA momentum map of Figure~\ref{fig:tddft_swa}(c), we find a pattern that is antisymmetric upon reflections at both the ($k_x,k_z$) and ($k_y,k_z$) planes, which can be found in combinations of the $|3,\pm 2\rangle$ spherical harmonics to the $f_{xyz}$ and $f_{z(x^2-y^2)}$ cubic harmonics (see Tables~\ref{tab_ang_xy_RH} and~\ref{tab_ang_xy_LH}).

We are now in a position to validate the predictions of the SWA model for the three Cartesian light incidence directions (field configurations $\ve A_1$,  $\ve A_2$, and $\ve A_3$) with the \emph{ab initio} TDDFT approach presented earlier. The resulting momentum maps are displayed in the right column of Figure~\ref{fig:tddft_swa} and, overall, we find good agreement between SWA and TDDFT regarding the principal symmetries in the CDAD patterns, i.e., the left-right antisymmetry for $\ve A_1$ [panel (a)], the up-down antisymmetry for $\ve A_2$ [panel (b)], as well as the left-right and up-down  antisymmetries for $\ve A_3$ [panel (c)]. For the latter case of normal incidence, the TDDFT map of the 4A HOMO is also very similar to the dichroism map of the 5A HOMO computed in a multiple scattering approach~\cite{Krueger2018}. In general, the TDDFT maps exhibit a richer structure compared to the SWA results, which we discuss in the following (see also Figure~\ref{fig:appx_linpol} for comparison). While the positions of the main lobes agree in both approaches, their exact shapes are different. This may be caused by the deviation of the scattering potentials from spherical symmetry, which is only taken into account in the TDDFT approach and may lead to the observed bending and stretching of the main emission lobes in the CDAD maps. Additionally, TDDFT predicts subfeatures located radially inside and outside the main lobes, which are not present in the SWA maps. Since a smaller $\ve k_{||}$ value is connected with a larger wavelength in real-space, one may associate these features with scattering effects of the photoelectrons in the potential of neighboring atoms. Such multiple-scattering effects are presently missing in our SWA approach but have been shown to be an important final-state effect in the photon energy dependence in graphene~\cite{Kern2023, Nozaki2024} or the circular dichroism in graphite~\cite{Krueger2022}. Despite these discrepancies, we conclude that the major effects leading to CDAD in organic molecules can indeed be captured by the SWA model, which means that the dichroism in such molecules is essentially an intraatomic effect. This makes clear why the claim by Wie{\ss}ner \emph{et al.}~\cite{Wiessner2014} cannot be true: The orbital phase these authors refer to is determined by the way in which constituent atomic orbitals (e.g., the $p_z$ orbitals of the frontier $\pi$ orbitals) are combined to form the molecular orbitals, while the CDAD, in this case being an intraatomic effect, does not know anything about this combination.

\begin{figure*}[htb]
  \includegraphics[width=13.6cm]{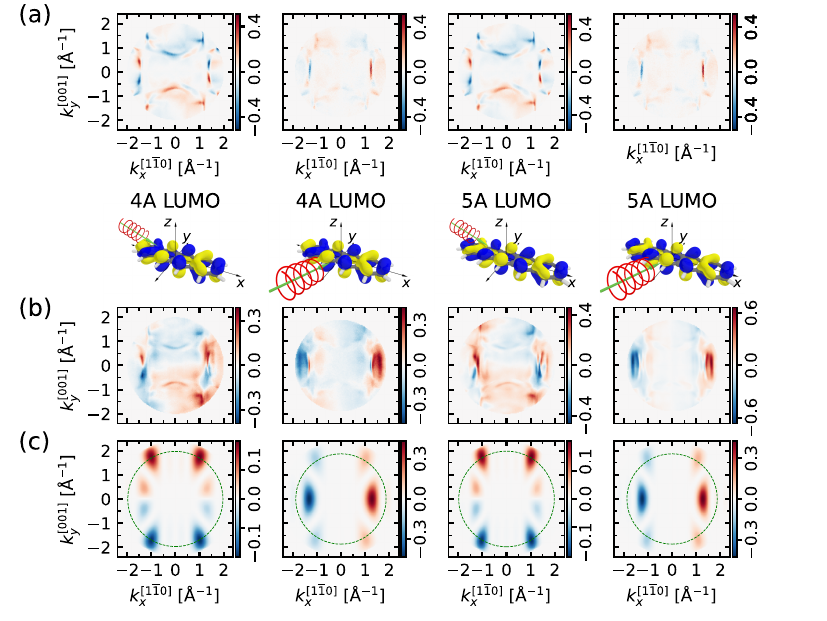}
  \caption{\label{fig:4A_5A_LUMO} CDAD maps of the LUMO in 4A and 5A for two different incidence directions of the circular polarized light as indicated in the sketches above panel~(b). (a) Experimental CDAD maps of the Cu(110) substrate at the same binding energies as the respective interface maps. (b) Experimental CDAD maps of 4A LUMO with the light incidence plane along the long and short molecular axis and 5A LUMO with the light incidence plane along the long and short molecular axis (from left to right) on Cu(110). (c) Maps of 4A and 5A, as simulated from SWA for isolated molecules, with corresponding experimental light polarization.}
\end{figure*}

\subsection{Results for the LUMOs of 4A and 5A}
The SWA allows us to compensate the shortcomings of the TDDFT t-SURFF method. While t-SURFF, being an \emph{ab initio} method, may in the best case exactly reproduce experimental results, it usually provides limited insight into the underlying physical processes. In contrast, the SWA model offers a clear physical interpretation. Additionally, t-SURFF simulations are computationally demanding, and TDDFT is most commonly applied to neutral isolated systems due to practical and numerical reasons. For 4A and 5A adsorbed on the Cu(110) surface, however, charge transfer from the surface populates the LUMO in both cases~\cite{Yang2019a,Ules2014}, which consequently becomes accessible for photoemission experiments, and shows in fact very rich CDAD patterns that are not accessible to TDDFT t-SURFF for neutral and isolated molecules.

Here, we focus on the CDAD patterns of the 4A and 5A LUMOs, both recorded at 0.85~eV below the Fermi level, respectively (see also the energy distribution curves plotted in Figure~\ref{fig:appx_edc}). The CDAD maps at those binding energies are presented in panel~(b) of Figure~\ref{fig:4A_5A_LUMO} with light polarization configurations sketched atop panel (b). Additionally, we show in panel (a) corresponding data for the bare substrate taken at the same binding energies. Turning to the case of the grazing incidence plane parallel to the $x$ axis first, we again find that the experimental maps in row~(b) for 4A (first column) and 5A (third column) are very similar. In comparison to the HOMO, the CDAD patterns from the LUMO are generally less pronounced and clearly dominated by the substrate, as can be seen from the corresponding CDAD maps of the bare substrate in panel~(a). This can be understood when consulting the SWA-simulated CDAD maps for the isolated molecules in the first and third columns of panel~(c): The features closest to the $k_y=0$ line appear much weaker than those around $k_y\approx\pm2$~$\textup{\AA}^{-1}$. These latter, stronger features are at the edge of the photoelectron horizon, indicated by the green circle, and therefore affected by fringe effects, making them difficult to measure. In contrast,  for the light incidence perpendicular to the long axis of the molecule the lobes centered on the $k_y=0$ line are well expressed in experiment. Again, the LUMO maps for 4A [Figure~\ref{fig:4A_5A_LUMO}~(b), second column] and 5A [Figure~\ref{fig:4A_5A_LUMO}~(b), fourth column] are very similar and also agree well with the SWA-simulated CDAD maps in panel~(c), respectively.

We regard the good agreement between the SWA and the experiments for the LUMO as noteworthy, because interactions with the substrate, such as hybridization and charge transfer, can potentially influence molecular frontier orbitals and may therefore also influence the CDAD. Clearly, this effect would not be accounted for by SWA calculations for isolated molecules. However, there are two arguments why such modifications are not expected to affect the CDAD strongly. First, standard POT studies with linearly polarized light of many molecular adsorbates have shown the frontier orbital structure itself to be remarkably robust, even when there is considerable charge transfer. Specifically, for the systems considered  here, comparing the photoemission momentum maps obtained using linearly polarized light with single-molecule calculations, we are assured that the orbital structure remains largely intact. As previously shown for 4A~\cite{Yang2019a} and 5A~\cite{Ules2014} on Cu(110), experimental momentum maps of frontier orbitals correspond closely to their theoretical counterparts from isolated molecules in the gas phase. Second, even if there were modifications of the molecular orbitals by the substrate, our SWA model offers an explanation why the CDAD is not greatly affected: On the one hand, such modifications are not expected to strongly influence the nature and selection of atomic wave functions from which the molecular orbitals are constructed. On the other hand, the good agreement between the SWA prediction and experiments shows that in systems studied here the CDAD is predominantly an intra-atomic effect in these constituent atomic orbitals.

Closer inspection of the results for both incidence directions reveals two further, interesting aspects. First, the molecular emission features located around the $k_y=0$ line exhibit a fine structure, unlike the corresponding lobes of the SWA-simulated patterns. This is caused by intermolecular band dispersion, as has been noted previously~\cite{Ules2014}. In fact, the filled LUMO of 5A/Cu(110) shows strong substrate-mediated  intermolecular dispersion with a band width of almost 1~eV~\cite{Ules2014} (cf.~the energy distribution curves in Figure~\ref{fig:appx_edc}), which appears even more pronounced in the CDAD maps, since here the subtraction of intensities is very sensitive to fine details. However, these dispersion effects, also present in Figure~\ref{fig:4A_5A_LUMO}(b) for 4A, do not alter the overall symmetry of the molecule-derived CDAD patterns, as the comparison with the SWA-simulated patterns in Figure~\ref{fig:4A_5A_LUMO}(c) shows. The second aspect becomes most evident through the comparison with CDAD maps for the clean substrate. Starting with the incidence plane parallel to the $x$ axis, we observe inversion symmetry (with respect to the $\Gamma$ point) of the substrate features at $k_x \approx \pm 1.4$~$\textup{\AA}^{-1}$ [Figure ~\ref{fig:4A_5A_LUMO}(a) first and third columns]. Remarkably, upon adsorption of the molecules, this symmetry is changed to the up-down antisymmetry of the molecular features [Figs.~\ref{fig:4A_5A_LUMO}(b) and~\ref{fig:4A_5A_LUMO}(c), first and third columns], which suggests that potentially existing hybrid molecule-substrate states---in this region of binding energy and momentum space---inherit their dichroic properties from the molecule, but not from the substrate. For the incidence plane parallel to the $y$ axis, the substrate maps are dominated by photoemission from the same band, this time however featuring a right-left antisymmetry [Figure ~\ref{fig:4A_5A_LUMO}(a) second and fourth columns]. Upon adsorption, the Cu $sp$ band remains visible (just touching the 4A molecular features on their sides facing the $\Gamma$ point), but with inverted contrast in comparison with the bare surface [Figure~\ref{fig:4A_5A_LUMO}(b), second column]. Conversely, for the 5A interface, the CDAD features from the substrate have almost vanished [Figure~\ref{fig:4A_5A_LUMO}(b), fourth column]. This interplay of substrate- and adsorbate-related circular dichroism for hybridized molecule-substrate states is remarkable and deserves further investigation; it is, however, beyond the capabilities of our simulations for isolated molecules and thus beyond the scope of the current work. 

\section{\label{sec:conclusions}Conclusions}
In this work, CDAD of photoelectrons from monolayers of uniaxially oriented tetracene (4A) and pentacene (5A) molecules on Cu(110) was investigated. The results for the HOMO orbitals for two perpendicular incidence planes of the circularly polarized light were found to be in remarkably good agreement with state-of-the-art TDDFT calculations for the isolated molecules. The similarity of the CDAD patterns of the two molecules, despite their HOMO orbitals having different symmetries, unambiguously demonstrates that the CDAD of photoelectrons from achiral molecules is not sensitive to the phase of the initial-state wave functions.

To gain insight into the mechanism that gives rise to the CDAD and simulate the observed LUMO patterns of the adsorbed molecules (neither of the two presently accessible to TDDFT), a simple final-state-scattering model was employed and found to adequately reproduce the principal features of the CDAD momentum space patterns.
This allowed us to study the CDAD effect in terms of dipole-allowed transitions between initial and final states of certain angular momentum in different light-field geometries. Our analysis highlights their importance and the necessity for concise theoretical modeling to understand the momentum-resolved circular dichroism in photoemission experiments.

As a major insight of our work, we demonstrate that the CDAD of achiral molecules is of \emph{intra}atomic nature and mainly arises from the orbital angular momentum of the atomic orbitals which constitute the frontier molecular orbitals. Thereby, we disprove previous claims that the CDAD can be used to extract phase information of the \emph{molecular} orbital. Furthermore, we showed that for certain experimental geometries, the CDAD can be sensitive to subtle details in the electronic structure. By controlling the incidence angle of the light with respect to the molecular orientation, it is therefore possible to extract minor angular-momentum components which contribute to the initial-state orbitals. For instance, normal light incidence is expected to map out the $d$ orbital content of a molecular orbital that is otherwise dominated by $p_z$ components. 
Moreover, the sensitivity of CDAD to the character of the initial state may also provide additional insights into the nature of hybrid states at molecule-metal interfaces. However, to apply the scattered-wave approximation to full adsorbate systems, an extension of the method to periodic systems is necessary.

\acknowledgements
We acknowledge Elettra Sincrotrone Trieste for providing access to its synchrotron radiation facilities. The computational results have been achieved using the HPC infrastructure of the University of Graz and the resources provided by Austrian Scientific Computing (ASC). This research was funded in in part by the Austrian Science Fund (FWF) (Project I3731). For the purpose of open access,the authors have applied a CC BY public copyright licence to any Author Accepted Manuscript version arising from this submission. Additional funding was provided from the European Union through the Synergy Grant Orbital Cinema (101071259) by the European Research Council (ERC). We acknowledge support from the Marie Sk{\l}odowska-Curie Doctoral Network TIMES (Grant No. 101118915) and SPARKLE (Grant No. 101169225), from the Italian Ministry of University and Research (MUR) under the PRIN 2022 Grant No 2022PX279E\_003, and from the Next Generation EU Partenariato Esteso NQSTI-Spoke 2 (THENCE-PE00000023).
This work was supported by the Cluster of Excellence Advanced Imaging of Matter (AIM) and Grupos Consolidados (IT1249-19). We acknowledge support from the Max Planck-New York City Center for Non-Equilibrium Quantum Phenomena. The Flatiron Institute is a division of the Simons Foundation

\section*{author contributions}
X.Y. conducted the experiment, together with G.Z., S.M., M.J., and V.F., supervised by S.S., M.G.R., and F.S.T. X.Y. and C.S.K. analyzed the experimental data, together with S.S. and M.G.R. C.S.K. conducted the TDDFT simulations, with the help of U.D.G. and A.R., and developed the SWA model, together with P.P. C.S.K. wrote the paper, together with F.S.T. and P.P. All authors contributed to the manuscript and the scientific discussion. U.D.G., A.R., F.S.T., and P.P. provided funding, and F.S.T. and P.P. were responsible for the overall project direction.

\section*{data availability}
All experimentally obtained datasets, as well as the input and output ﬁles necessary to reproduce the simulations, are available from the J{\"u}lich data repository~\cite{JuelichData}.

\appendix

\section{\label{sec:appx_exp}Experimental details}
\begin{figure}[htb]
  \includegraphics[width=6.cm]{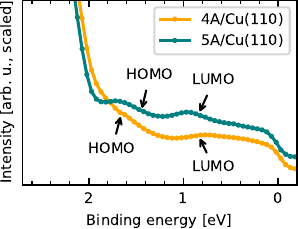}
  \caption{\label{fig:appx_edc}Energy distribution curves for 4A and 5A on Cu(110) with binding energy measured with respect to the Fermi level. Energy positions for HOMO and LUMO maps are marked with arrows.}
\end{figure}

Photoemission experiments were conducted at the NanoESCA beamline~\cite{Schneider2012} of Elettra Sincrotrone Trieste (Italy), utilizing a photoemission electron microscope from FOCUS GmbH. 4A and 5A were sublimated in vacuum and adsorbed onto a Cu(110) surface, which was prepared through several cycles of Ar$^+$-ion sputtering and annealing at 550$^{\circ}$C. At room temperature, both oligoacenes formed well-ordered monolayers with flat-lying molecules aligned along the [1$\overline{1}$0] direction of Cu, as confirmed by low-energy electron diffraction and photoemission tomography using linearly polarized ultraviolet light~\cite{Yang2019a}. For CDAD measurements, right- and left-handed circularly polarized light with a photon energy of 35~eV was used. At an oblique incidence of 65$^{\circ}$ relative to the surface normal, the light was directed either along the [001] or [1$\overline{1}$0] directions of copper, i.e., either perpendicular to or along the long molecular axis. Photoemission momentum maps were recorded at the kinetic energy of photoelectrons corresponding to the HOMOs and LUMOs of 4A and 5A~\cite{Yang2019a}. To minimize potential beam damage from the focused synchrotron radiation, the sample was continuously moved laterally during measurements. After averaging and centering, CDAD maps were derived by evaluating Eq.~\ref{eq:cdad_definition} for the experimental maps. Figure~\ref{fig:appx_edc} shows the momentum-integrated EDCs for the 4A and 5A monolayers. Within the broad LUMO and HOMO regions, the energy positions of the HOMO maps of Figure~\ref{fig:4A_5A_HOMO} and the LUMO maps of Figure~\ref{fig:4A_5A_LUMO} are marked by arrows.

\section{\label{sec:appx_tddft}Details on TDDFT simulations}
For the simulations of the isolated 4A and 5A molecules with the Octopus code, we used a grid spacing of 0.2~$\textup{\AA}$ and a spherical simulation box with 40~$\textup{\AA}$ radius. We propagated the system for 18~fs under the influence of a 35~eV (30~eV in the case of Figure~\ref{fig:tddft_swa}) monochromatic pulse with a $\sin^2$ hull that reached its maximum field strength of $6\times 10^{-5}$~a.u. after 8~fs and is zero after 16~fs. The maximum field strength corresponds to an intensity of $1.25\times 10^8$~W/cm$^2$. Throughout the simulation time, we recorded the flux of photoelectrons through a spherical detector surface placed at 20~$\textup{\AA}$ radius and after which a complex absorbing potential of $-0.2$~a.u. was applied in order to avoid electrons from reflecting at the boundaries of the simulation box~\cite{DeGiovannini2015}.

\begin{figure}[htb]
  \includegraphics[width=8.4cm]{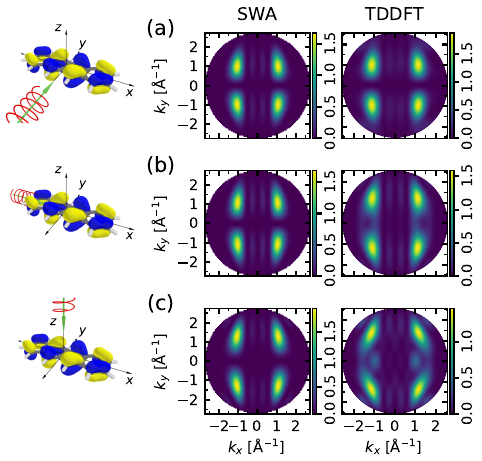}
  \caption{\label{fig:appx_linpol}Same as Figure~\ref{fig:tddft_swa} albeit with $\mathcal{I}_{\mathrm{RH}} + \mathcal{I}_{\mathrm{LH}}$ instead of $\mathcal{I}_{\mathrm{RH}} - \mathcal{I}_{\mathrm{LH}}$, which corresponds to linear polarized light along the three cartesian axes, respectively.}
\end{figure}

\section{\label{sec:appx_me}Details on the matrix element}
\begin{figure}[htb]
  \includegraphics[width=7.cm]{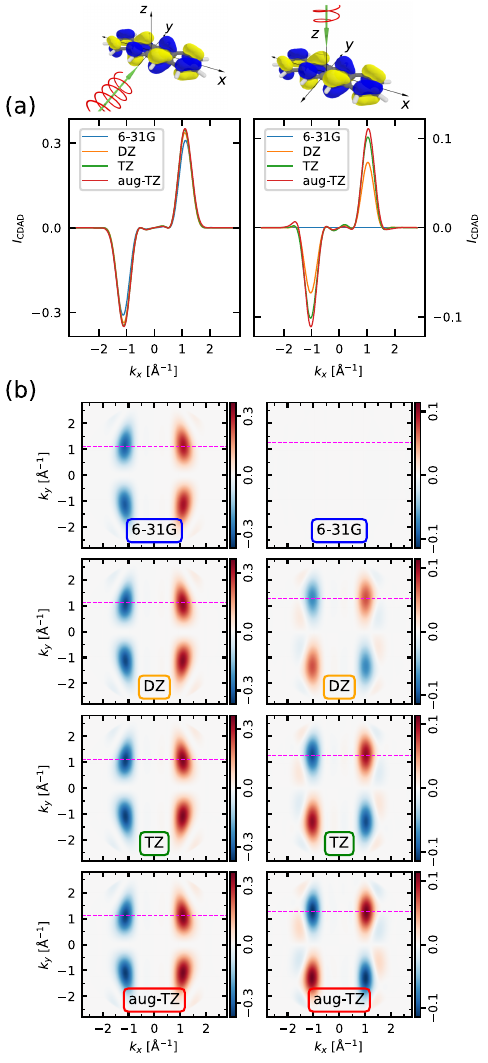}
  \caption{\label{fig:appx_basis_sets}Convergence of CDAD maps with different basis sets in the SWA. In panel (b), we show momentum maps for different basis sets (see insets) and in panel (a) line plots for the path in momentum space that is indicated by the dashed lines in the maps of panel (b). Left column: results for field polarization $\widehat{\ve A}_1=(\pm \mathrm i \hat{\ve x} + \hat{\ve z})/\sqrt{2}$, right column: results for field polarization $\widehat{\ve A}_3=(\hat{\ve x} \mp \mathrm i \hat{\ve y})/\sqrt{2}$. For each column, all maps are normalized to the same maximum and minimum values to allow for better comparison.}
\end{figure}

With the definitions $\Psi (\ve r) = \langle \ve r | i \rangle$ and $\chi_{\ve k} (\ve r) = \langle \ve r | f_{\ve k} \rangle$, the matrix element $\mathcal{M}(\ve k) = \langle f_{\ve k} | \ve A \cdot \nabla | i \rangle$ in Eq.~\ref{eq:me} can be written as an integral in real space. Making use of the IACA, we can split the final state into a sum over atomic contributions,
\begin{align}
  \label{eq:fs_decompose}
  \chi_{\ve k} (\ve r) = \sum_a \mathrm{e}^{\mathrm i \ve k \ve R_a}\widetilde{\chi}_{\ve k} (\ve r_a),
\end{align}
which allows us to also consider the matrix element of one such atomic contribution:
\begin{align}
  \label{eq:me_pw}
  \mathcal{M}_a(\ve k) &= \int \mathrm d^3 \ve{r}_a \;
  \widetilde{\chi}_{\ve k}^*(\ve r_a) \ve{A} \cdot \nabla_a \Psi(\ve r_a).
\end{align}
Next, we evaluate the action of the gradient operator on functions of well-defined angular momentum~\cite{Arfken2013_16}:
\begin{align}
  \label{eq:gradient_initial}
  \nabla_a \Psi(\ve r_a) =
  \sum_{n l m} c_{n l m}^a \Bigg[&\left( \frac{l}{2l+1}\right)^\frac{1}{2}
  \left(\frac{\partial}{\partial r_a} +\frac{l+1}{r_a}\right)\nonumber \\
  & R_{n l}(r_a)\mathbf{\mathcal{Y}}_{l, l-1, m}(\widehat{\ve r}_a)
  \nonumber \\
  -&\left( \frac{l+1}{2l+1}\right)^\frac{1}{2} 
  \left(\frac{\partial}{\partial r_a} -\frac{l}{r_a}\right) \nonumber \\
  & R_{n l}(r_a)\mathbf{\mathcal{Y}}_{l, l+1, m}(\widehat{\ve r}_a)
  \Bigg].
\end{align}
The vector spherical harmonics $\mathbf{\mathcal{Y}}_{J, L, M}$ are defined as
\begin{align}
  \mathbf{\mathcal{Y}}_{J, L, M} (\widehat{ \ve r}) =
  \sum_{m=-L}^L \sum_{\mu=-1}^1 c_{\mathrm{CG}}(L,1,J;m,\mu,M)
  \mathrm{Y}_{L m}(\widehat{\ve r}) \boldsymbol{\xi}_{\mu},
\end{align}
with the Clebsch-Gordan coefficients $c_{\mathrm{CG}}$ and 
\begin{align}
  \label{eq:xi}
  \boldsymbol{\xi}_{-1} = \frac{1}{\sqrt 2} \left( \hat{\ve x} - \mathrm i \hat{\ve y}\right), \;\;
  \boldsymbol{\xi}_{0} = \hat{\ve z}, \;\;
  \boldsymbol{\xi}_{1} = -\frac{1}{\sqrt 2} \left( \hat{\ve x} + \mathrm i \hat{\ve y}\right).
\end{align}
Here, we use the Condon-Shortley phase convention~\cite{Condon1935} and understand $c_{\mathrm{CG}}(j_1, j_2, J; m_1, m_2, M)$ in the sense that $|j_1, m_1\rangle$ and $|j_2, m_2\rangle$ are coupled to form the total angular momentum state $|J, M\rangle$~\cite{Arfken2013_16}.

Introducing abbreviations for the radial parts
\begin{align}
  \Delta_{n}^{(l+1)}(r_a) \equiv
  -\left( \frac{l+1}{2l+1}\right)^\frac{1}{2} 
  \left(\frac{\partial}{\partial r_a} -\frac{l}{r_a}\right) R_{n l}(r_a),
\end{align}
as well as
\begin{align}
  \Delta_{n}^{(l-1)}(r_a ) \equiv 
  \left( \frac{l}{2l+1}\right)^\frac{1}{2}
    \left(\frac{\partial}{\partial r_a} +\frac{l+1}{r_a}\right) R_{n l}(r_a),
\end{align}
we can write Eq.~\ref{eq:gradient_initial} in a more compact way as
\begin{align}
  \label{eq:gradient_initial_compact}
  \nabla \Psi(\ve r_a) =
  \sum_{n l m} c_{n l m}^a \Big[&\Delta_{n}^{(l-1)}(r_a)\mathbf{\mathcal{Y}}_{l, l-1, m}(\widehat{\ve r}_a) \nonumber \\
  +&\Delta_{n}^{(l+1)}(r_a )\mathbf{\mathcal{Y}}_{l, l+1, m}(\widehat{\ve r}_a)\Big].
\end{align}
Evidently, the action of the gradient operator on an initial state of well-defined angular momentum $l$ thus results in \emph{two} states with angular momentum $l+1$ and $l-1$.

Incorporating the final state $\widetilde{\chi}(\ve r_a)$, as defined by Eqs.~\ref{eq:final_state} and~\ref{eq:fs_decompose}, we can write the matrix element (at site $a$ and without continuum normalization factors) as
\begin{align}
  \label{eq:me_sw}
  \mathcal{M}_a(\ve k) = 4\pi \sum_{l'm'}\int &\mathrm d^3 \ve{r}_a \;
  \mathrm{e}^{-\mathrm i \sigma_{l'}^Z} j_{l'} (k r_a)
  (-\mathrm i)^{l'}\mathrm{Y}_{l' m'}(\widehat{\ve k})  \nonumber \\
  &\mathrm{Y}_{l' m'}^*(\widehat{\ve r}_a) \ve A \cdot
  \nabla_a \Psi(\ve r_a).
\end{align}
Next, we insert Eq.~\ref{eq:gradient_initial_compact} into the matrix element of Eq.~\ref{eq:me_sw} and obtain
\begin{align}
  \label{eq:me_a_sep_pre}
  \mathcal{M}_a(\ve k) = 4\pi \sum_{l' m'} \sum_{nlm} c_{n l m}^a &\mathrm{e}^{-\mathrm i \sigma_{l'}^Z}
  \int \mathrm d \Omega_a \mathrm{Y}_{l' m'}(\widehat{\ve k}) 
  \mathrm{Y}_{l' m'}^*(\widehat{\ve r}_a) \nonumber \\  
  \ve A\cdot\Big[&\mathcal{R}_{a,n,l'}^{(l-1)}
  \mathbf{\mathcal{Y}}_{l, l-1, m}(\widehat{\ve r}_a) \nonumber \\
  +&\mathcal{R}_{a,n,l'}^{(l+1)}
  \mathbf{\mathcal{Y}}_{l, l+1, m}(\widehat{\ve r}_a)\Big],
\end{align}
where we have defined the radial integrals as
\begin{align}
  \label{eq:def_rad_me}
  \mathcal{R}_{a,n,l'}^{(l \pm 1)}\equiv (-\mathrm i)^{l'}
  \int \mathrm d r_{a}\; r_{a}^2\, j_{l'}(k r_a) \Delta_{n}^{(l \pm 1)}(r_a).
\end{align}
For each dipole-allowed angular momentum channel $l+1$ and $l-1$, we can furthermore isolate the angular integrals $\mathcal{T}_{m,l',m'}^{(l,l \pm 1)}$ and write the matrix element as
\begin{align}
  \label{eq:me_a_sep}
  \mathcal{M}_a(\ve k) = 4\pi A_0 \sum_{l' m'} \sum_{nlm} &c_{n l m}^a \mathrm{e}^{-\mathrm i \sigma_{l'}^Z}
  \Big[\mathcal{R}_{a,n,l'}^{(l-1)}
  \mathcal{T}_{l',m,m'}^{(l,l-1)}(\widehat{\ve A}) \nonumber \\
  &+\mathcal{R}_{a,n,l'}^{(l+1)}
  \mathcal{T}_{l',m,m'}^{(l,l+1)}(\widehat{\ve A})\Big].
\end{align}
The angular integrals are functions of the normalized field configuration $\widehat{\ve A}$ (using $\ve A = A_0 \widehat{\ve A}$):
\begin{align}
  \mathcal{T}_{l',m,m'}^{(l,l\pm 1)}(\widehat{\ve A}) \equiv
  &\int \mathrm d \Omega_a \mathrm{Y}_{l' m'}(\widehat{\ve k})\mathrm{Y}_{l' m'}^*(\widehat{\ve r}_a) \widehat{\ve A}\cdot
  \mathbf{\mathcal{Y}}_{l, l \pm 1, m}(\widehat{\ve r}_a) \nonumber \\
  = &\sum_{m''=-(l\pm 1)}^{(l\pm 1)} \mathrm{Y}_{l' m'}(\widehat{\ve k}) \nonumber \\
  &\sum_{\mu=-1}^1  c_{\mathrm{CG}}(l\pm 1,1,l;m'',\mu,m)\,
  \widehat{\ve A} \cdot \boldsymbol{\xi}_{\mu} \nonumber \\
  &\int \mathrm d \Omega_a \mathrm{Y}_{(l \pm 1) m''}(\widehat{\ve r}_a)
   \mathrm{Y}_{l' m'}^*(\widehat{\ve r}_a).
\end{align}
The orthogonality relations of the spherical harmonics fix $l'=l \pm 1$, $m''=m'$. This  simplifies the angular integrals to
\begin{align}
  \label{eq:ang_me_red}
  \mathcal{T}_{m,m'}^{(l,l\pm 1)}(\widehat{\ve A}) =
  &\sum_{m''=-(l\pm 1)}^{(l\pm 1)} \delta_{m'' m'}\mathrm{Y}_{(l\pm 1) m''}(\widehat{\ve k}) \alpha_{m,m''}^{(l,l\pm 1)}(\widehat{\ve A}),
\end{align}
where we have defined the angular coefficients as
\begin{align}
  \label{eq:def_ang_coef}
  \alpha_{m,m''}^{(l,l\pm 1)}(\widehat{\ve A}) \equiv
  \sum_{\mu=-1}^1 c_{\mathrm{CG}}(l\pm 1,1,l;m'',\mu,m)\,\widehat{\ve A}\cdot \boldsymbol{\xi}_{\mu}.
\end{align}
Inserting Eq.~\ref{eq:ang_me_red} into Eq.~\ref{eq:me_a_sep} then finally leads to the matrix element
\begin{align}
  \label{eq:me_a_}
  \mathcal{M}_a(\ve k) = 4 &\pi A_0 \sum_{nlm} c_{n l m}^a \\
  \Bigg[&\mathrm{e}^{-\mathrm i \sigma_{l-1}^Z}\mathcal{R}_{a,n}^{(l-1)}
  \hspace{-3mm}\sum_{m'=-(l-1)}^{l-1}\hspace{-4mm}\mathrm{Y}_{(l-1)\, m'}(\widehat{\ve k})
  \alpha_{m,m'}^{(l,l- 1)}(\widehat{\ve A}) \nonumber \\
  +
  &\mathrm{e}^{-\mathrm i \sigma_{l+1}^Z}\mathcal{R}_{a,n}^{(l+1)}\hspace{-2mm}
  \sum_{m'=-(l+1)}^{l+1}\hspace{-4mm}\mathrm{Y}_{(l+1)\, m'}(\widehat{\ve k})
  \alpha_{m,m'}^{(l,l+ 1)}(\widehat{\ve A})\Bigg], \nonumber
\end{align}
where we have dropped the now-obsolete $l'=l \pm 1$ indices in $\mathcal{R}_{a,n}^{(l\pm 1)}$ and $\delta_{m' m''}$ has been evaluated. Finally, when summing over the contributions from each atom (located at $\ve R_a$), we have to take Eq.~\ref{eq:fs_decompose} into account~\cite{Grobman1978}, such that
\begin{align}
  \label{eq:me_IACA}
  \mathcal{M}(\ve k) = \sum_a \mathrm{e}^{\mathrm i \ve k \ve R_a} \mathcal{M}_a(\ve k).
\end{align}

\section{\label{sec:appx_contributions}Different contributions to the CDAD signal}
In order to decompose the CDAD signal into intra- and interatomic contributions~\cite{Sidilkover2025}, we write Eq.~\ref{eq:me_a_} in a more compact way:
\begin{align}
  \mathcal{M}_a(\ve k) = 4 \pi A_0 \sum_{nlm} c_{n l m}^a
  F_{a,n l m}(\ve k; \widehat{\ve A}, Z).
\end{align}
Here, we have defined the on-site matrix element $F_{a,n l m}(\ve k; \widehat{\ve A}, Z)$ as the term inside the square brackets of Eq.~\ref{eq:me_a_}. Summing over the contributions from all sites as in Eq.~\ref{eq:me_IACA}, we get
\begin{align}
  \mathcal{M}(\ve k) = 4 \pi A_0 \sum_a \sum_{nlm} c_{n l m}^a
  \mathrm{e}^{\mathrm i \ve k \ve R_a} F_{a,n l m}(\ve k; \widehat{\ve A}, Z).
\end{align}
When evaluating the CDAD signal as the difference signal between photoemission intensities with two different circular polarizations, $\ve A^{(+)}$ and $\ve A^{(-)}$, and compressing the initial state quantum numbers to $j=\{n l m\}$, we obtain
\begin{align}
  \label{eq:intensity_D}
  \mathcal{I}_{\mathrm{CDAD}} = &\bigg|\mathcal{M}^{(+)}\bigg|^2 - \bigg|\mathcal{M}^{(-)}\bigg|^2=\\ \nonumber
  &(4 \pi A_0)^2\sum_{a, a'} \sum_{j, j'}
  c_j^a (c_{j'}^{a'})^* \mathrm{e}^{\mathrm i \ve k (\ve R_a-\ve R_{a'})} \\ \nonumber
   &\bigg(F_{a,j}(\ve k; \widehat{\ve A}^{(+)}, Z)F_{a',j'}^*(\ve k; \widehat{\ve A}^{(+)}, Z)+ \\ \nonumber
   &-F_{a,j}(\ve k; \widehat{\ve A}^{(-)}, Z)F_{a',j'}^*(\ve k; \widehat{\ve A}^{(-)}, Z) \bigg).
\end{align}
In the spirit of the discussion following Eq.~\ref{eq:cdad_xz} in the main text, we now assume photoemission from an initial state, to which only a single atomic orbital (quantum number $j$) per site $a$ contributes. This could be, for instance, single $p_z$ orbitals located at each atomic site, as in the H{\"u}ckel model for organic molecules. Then the sums over $j,j'$ collapse and, importantly,
only a single matrix element $F_{a,j}(\ve k; \widehat{\ve A}^{(\pm)}, Z)\equiv F(\ve k; \widehat{\ve A}^{(\pm)}, Z)$ will appear in Eq.~\ref{eq:intensity_D} [the radial integrals $R_{a,n}^{(l\pm 1)}$ in Eq.~\ref{eq:me_a_} will be the same for all values of $a$] and can therefore be factored out. Defining $\Delta F=F(\ve k; \widehat{\ve A}^{(+)}, Z)F^*(\ve k; \widehat{\ve A}^{(+)}, Z)-F(\ve k; \widehat{\ve A}^{(-)}, Z)F^*(\ve k; \widehat{\ve A}^{(-)}, Z)$, we thus get
\begin{align}
  \mathcal{I}_{\mathrm{CDAD}}^{(j)} = (&4 \pi A_0)^2\Delta F\sum_{a, a'}
  c_j^a (c_j^{a'})^* \mathrm{e}^{\mathrm i \ve k (\ve R_a-\ve R_{a'})}.
\end{align}
The double sum in this term can be split into intraatomic contributions, where $a=a'$, and interatomic contributions, where $a\neq a'$, respectively:
\begin{align}
  \mathcal{I}_{\mathrm{CDAD}}^{(j)} = (4 \pi A_0)^2\Delta F\bigg[ &\sum_{a} \left|c_j^{a} \right|^2\\ \nonumber
  &+\sum_{a \neq a'}
  c_j^a (c_j^{a'})^* \mathrm{e}^{\mathrm i \ve k (\ve R_a-\ve R_{a'})} \bigg].
\end{align}
Finally, using the fact that $(\ve R_a - \ve R_{a'})=-(\ve R_{a'} - \ve R_a)$, we can simplify this to a real-valued expression, as demanded for an intensity signal:
\begin{align}
  \mathcal{I}_{\mathrm{CDAD}}^{(j)} = (4 \pi A_0)^2\Delta F\bigg[&\sum_{a} \left|c_j^{a} \right|^2 \\ \nonumber
  2 +&\sum_{a < a'}
  c_j^a (c_j^{a'})^* \cos\left[ \ve k (\ve R_a-\ve R_{a'})\right] \bigg].
\end{align}

\section{\label{appx:ang_tables}Additional angular matrix elements}
\begin{table}[htb]
\caption{\label{tab_ang_yz}Angular coefficients $\alpha_{m,m'}^{(l,l\pm 1)}(\widehat{\ve A}_2)$ for light polarization $\widehat{\ve A}_2=(\mp \mathrm i \hat{\ve y}+\hat{\ve z})/\sqrt{2}$ for combinations of initial states $|i\rangle=|l,m\rangle $ (columns) and final states $|f\rangle=|l\pm1,m'\rangle$ (rows) for both dipole-allowed channels. The $\pm$ sign reflects a sign change upon change of helicity.}
\begin{center} 
\setlength{\tabcolsep}{2pt}
\renewcommand{\arraystretch}{1.5}
\begin{tabular}{l|l|l|l|l|l|}
\cline{2-6}
& \diagbox{$|f\rangle$}{$|i\rangle$} & $|0,0\rangle$ & $|1,-1\rangle$ & $|1,0\rangle$ & $|1,1\rangle$\\ \cline{2-6}
\vspace{-.2cm} \\
$l-1$ \\
\hline \vspace{-.5cm} \\
\hline
 & $| 0 , 0 \rangle$ &    & $ \pm \frac{1}{2} $ & $ \frac{\sqrt{2}}{2} $ & $ \pm \frac{1}{2} $ \\ \cline{2-6}
\vspace{-.2cm} \\
$l+1$ \\
\hline \vspace{-.5cm} \\
\hline
 & $| 1 , -1 \rangle$ & $ \pm \frac{\sqrt{3}}{6} $ &    &    &    \\ \cline{2-6}
 & $| 1 , 0 \rangle$ & $ - \frac{\sqrt{6}}{6} $ &    &    &    \\ \cline{2-6}
 & $| 1 , 1 \rangle$ & $ \pm \frac{\sqrt{3}}{6} $ &    &    &    \\ \cline{2-6}
 & $| 2 , -2 \rangle$ &    & $ \pm \frac{\sqrt{15}}{10} $ &    &    \\ \cline{2-6}
 & $| 2 , -1 \rangle$ &    & $ - \frac{\sqrt{15}}{10} $ & $ \pm \frac{\sqrt{30}}{20} $ &    \\ \cline{2-6}
 & $| 2 , 0 \rangle$ &    & $ \pm \frac{\sqrt{10}}{20} $ & $ - \frac{\sqrt{5}}{5} $ & $ \pm \frac{\sqrt{10}}{20} $ \\ \cline{2-6}
 & $| 2 , 1 \rangle$ &    &    & $ \pm \frac{\sqrt{30}}{20} $ & $ - \frac{\sqrt{15}}{10} $ \\ \cline{2-6}
 & $| 2 , 2 \rangle$ &    &    &    & $ \pm \frac{\sqrt{15}}{10} $ \\ \cline{2-6}
\cline{2-6}
\end{tabular}
\end{center}
\end{table}

\begin{table*}[htb]
\caption{\label{tab_ang_xy_RH}Angular coefficients $\alpha_{m,m'}^{(l,l\pm 1)}(\widehat{\ve A}_3^{(RH)})$ with light polarization $\widehat{\ve A}_3^{(RH)}=(\hat{\ve x}- \mathrm i\hat{\ve y})/\sqrt{2}$ for combinations of initial-state (columns) and final-state (rows) angular momentum quantum numbers represented as $|l,m\rangle$ for both dipole-allowed $(l \pm 1)$ channels.}
\begin{center} 
\setlength{\tabcolsep}{2pt}
\renewcommand{\arraystretch}{1.5}
\begin{tabular}{l|l|l|l|l|l|l|l|l|l|l|}
  \cline{2-11}
& \diagbox{$|f\rangle$}{$|i\rangle$} & $|0,0\rangle$ & $|1,-1\rangle$ & $|1,0\rangle$ & $|1,1\rangle$ & $|2,-2\rangle$ & $|2,-1\rangle$ & $|2,0\rangle$ & $|2,1\rangle$ & $|2,2\rangle$\\ \cline{2-11}
\vspace{-.2cm} \\
$l-1$ \\
\hline \vspace{-.5cm} \\
\hline
 & $| 0 , 0 \rangle$ &    & $ 1 $ &  &  &    &    &    &    &    \\ \cline{2-11}
 & $| 1 , -1 \rangle$ &    &    &    &    & $ 1 $ &  &  &    &    \\ \cline{2-11}
 & $| 1 , 0 \rangle$ &    &    &    &    &    & $ \frac{\sqrt{2}}{2} $ &  &  &    \\ \cline{2-11}
 & $| 1 , 1 \rangle$ &    &    &    &    &    &    & $ \frac{\sqrt{6}}{6} $ &  &  \\ \cline{2-11}
\vspace{-.2cm} \\
$l+1$ \\
\hline \vspace{-.5cm} \\
\hline
 & $| 1 , 1 \rangle$ & $ \frac{\sqrt{3}}{3} $ &    &    &    &    &    &    &    &    \\ \cline{2-11}
 & $| 2 , -2 \rangle$ &    &  &    &    &    &    &    &    &    \\ \cline{2-11}
 & $| 2 , -1 \rangle$ &    &  &  &    &    &    &    &    &    \\ \cline{2-11}
 & $| 2 , 0 \rangle$ &    & $ \frac{\sqrt{10}}{10} $ &  &  &    &    &    &    &    \\ \cline{2-11}
 & $| 2 , 1 \rangle$ &    &    & $ \frac{\sqrt{30}}{10} $ &  &    &    &    &    &    \\ \cline{2-11}
 & $| 2 , 2 \rangle$ &    &    &    & $ \frac{\sqrt{15}}{5} $ &    &    &    &    &    \\ \cline{2-11}
 & $| 3 , -3 \rangle$ &    &    &    &    &  &    &    &    &    \\ \cline{2-11}
 & $| 3 , -2 \rangle$ &    &    &    &    &  &  &    &    &    \\ \cline{2-11}
 & $| 3 , -1 \rangle$ &    &    &    &    & $ \frac{\sqrt{21}}{21} $ &  &  &    &    \\ \cline{2-11}
 & $| 3 , 0 \rangle$ &    &    &    &    &    & $ \frac{\sqrt{7}}{7} $ &  &  &    \\ \cline{2-11}
 & $| 3 , 1 \rangle$ &    &    &    &    &    &    & $ \frac{\sqrt{14}}{7} $ &  &  \\ \cline{2-11}
 & $| 3 , 2 \rangle$ &    &    &    &    &    &    &    & $ \frac{\sqrt{210}}{21} $ &  \\ \cline{2-11}
 & $| 3 , 3 \rangle$ &    &    &    &    &    &    &    &    & $ \frac{\sqrt{35}}{7} $ \\ \cline{2-11}
\end{tabular}
\end{center}
\end{table*}

\begin{table*}[htb]
\caption{\label{tab_ang_xy_LH}Angular coefficients $\alpha_{m,m'}^{(l,l\pm 1)}(\widehat{\ve A}_3^{(LH)})$ with light polarization $\widehat{\ve A}_3^{(LH)}=(\hat{\ve x}+ \mathrm i\hat{\ve y})/\sqrt{2}$ for combinations of initial-state (columns) and final-state (rows) angular momentum quantum numbers represented as $|l,m\rangle$ for both dipole-allowed channels.}
\begin{center} 
\setlength{\tabcolsep}{2pt}
\renewcommand{\arraystretch}{1.5}
\begin{tabular}{l|l|l|l|l|l|l|l|l|l|l|}
\cline{2-11}
& \diagbox{$|f\rangle$}{$|i\rangle$} & $|0,0\rangle$ & $|1,-1\rangle$ & $|1,0\rangle$ & $|1,1\rangle$ & $|2,-2\rangle$ & $|2,-1\rangle$ & $|2,0\rangle$ & $|2,1\rangle$ & $|2,2\rangle$\\ \cline{2-11}
\vspace{-.2cm} \\
$l-1$ \\
\hline \vspace{-.5cm} \\
\hline
 & $| 0 , 0 \rangle$ &    &    &    & $ -1 $ &    &    &    &    &    \\ \cline{2-11}
 & $| 1 , -1 \rangle$ &    &    &    &    &    &    & $ - \frac{\sqrt{6}}{6} $ &    &    \\ \cline{2-11}
 & $| 1 , 0 \rangle$ &    &    &    &    &    &    &    & $ - \frac{\sqrt{2}}{2} $ &    \\ \cline{2-11}
 & $| 1 , 1 \rangle$ &    &    &    &    &    &    &    &    & $ -1 $ \\ \cline{2-11}
\vspace{-.2cm} \\
$l+1$ \\
\hline \vspace{-.5cm} \\
\hline
 & $| 1 , -1 \rangle$ & $ - \frac{\sqrt{3}}{3} $ &    &    &    &    &    &    &    &    \\ \cline{2-11}
 & $| 1 , 0 \rangle$ &    &    &    &    &    &    &    &    &    \\ \cline{2-11}
 & $| 1 , 1 \rangle$ &    &    &    &    &    &    &    &    &    \\ \cline{2-11}
 & $| 2 , -2 \rangle$ &    & $ - \frac{\sqrt{15}}{5} $ &    &    &    &    &    &    &    \\ \cline{2-11}
 & $| 2 , -1 \rangle$ &    &    & $ - \frac{\sqrt{30}}{10} $ &    &    &    &    &    &    \\ \cline{2-11}
 & $| 2 , 0 \rangle$ &    &    &    & $ - \frac{\sqrt{10}}{10} $ &    &    &    &    &    \\ \cline{2-11}
 & $| 2 , 1 \rangle$ &    &    &    &    &    &    &    &    &    \\ \cline{2-11}
 & $| 2 , 2 \rangle$ &    &    &    &    &    &    &    &    &    \\ \cline{2-11}
 & $| 3 , -3 \rangle$ &    &    &    &    & $ - \frac{\sqrt{35}}{7} $ &    &    &    &    \\ \cline{2-11}
 & $| 3 , -2 \rangle$ &    &    &    &    &    & $ - \frac{\sqrt{210}}{21} $ &    &    &    \\ \cline{2-11}
 & $| 3 , -1 \rangle$ &    &    &    &    &    &    & $ - \frac{\sqrt{14}}{7} $ &    &    \\ \cline{2-11}
 & $| 3 , 0 \rangle$ &    &    &    &    &    &    &    & $ - \frac{\sqrt{7}}{7} $ &    \\ \cline{2-11}
 & $| 3 , 1 \rangle$ &    &    &    &    &    &    &    &    & $ - \frac{\sqrt{21}}{21} $ \\ \cline{2-11}
\end{tabular}
\end{center}
\end{table*}

\bibliography{bibliography}
\end{document}